# DESIGN OF PARITY PRESERVING LOGIC BASED FAULT TOLERANT REVERSIBLE ARITHMETIC LOGIC UNIT

Rakshith Saligram[1]   Shrihari Shridhar Hegde[1]   Shashidhar A Kulkarni[1]

H.R.Bhagyalakshmi[1]  and M.K. Venkatesha[2]

[1]Department of Electronics and Communication, BMS College of Engineering, Visvesvaraya Technological University, Bangalore, India

[2]Department of Electronics and Communication, RNS Institute of Technology, Visvesvaraya Technological University, Bangalore, India
.

## ABSTRACT

*Reversible Logic is gaining significant consideration as the potential logic design style for implementation in modern nanotechnology and quantum computing with minimal impact on physical entropy .Fault Tolerant reversible logic is one class of reversible logic that maintain the parity of the input and the outputs. Significant contributions have been made in the literature towards the design of fault tolerant reversible logic gate structures and arithmetic units, however, there are not many efforts directed towards the design of fault tolerant reversible ALUs. Arithmetic Logic Unit (ALU) is the prime performing unit in any computing device and it has to be made fault tolerant. In this paper we aim to design one such fault tolerant reversible ALU that is constructed using parity preserving reversible logic gates. The designed ALU can generate up to seven Arithmetic operations and four logical operations.*

## 1. INTRODUCTION

Researchers like Landauer [1] and Bennett [2] have shown that every bit of information lost will generate kTlog2 joules of energy, whereas the energy dissipation would not occur, if a computation is carried out in a reversible way. k is Boltzmann's constant and T is absolute temperature at which computation is performed. Thus reversible circuits will be the most important one of the solutions of heat dissipation in future circuit design. Reversible computing is motivated by the Von Neumann Landauer (VNL) principle, (a theorem of modern physics telling us that ordinary irreversible logic operation which destructively overwrite previous outputs incur a fundamental minimum energy cost). Such operations typically dissipate roughly the logic signal energy, itself irreducible due to thermal noise. This fact threatens to end improvements in practical computer performance within the next few decades. However, computers based mainly on reversible logic operations can reuse a fraction of the signal energy that theoretically can approach arbitrarily near to 100% as the quality of the hardware is improved, reopening the door to arbitrarily high computer performance at a given level of power dissipation.

The advancement in VLSI designs, portable device technologies and increasingly high computation requirements, lead to the circuit design of faster, smaller and more complex electronic systems at the expense of lots of heat dissipation which would reduce the life of the circuit. Thus power consumption becomes an important issue in modern design. The power dissipation that is tolerable in a given application context is always limited by some practical consideration, such as a requirement that a limited supply of available energy (such as in a battery) not be used up within a given time, or by the limited rate of heat removal in one's cooling system, or by a limited operating budget available for buying energy. Thus, improving system performance generally requires increasing the average energy efficiency of useful operations. It has been clearly demonstrated by Frank [17] that reversible computing is the only viable option to overcome the power dissipation. The primary motivation for reversible computing lies in the fact that it provides the only way (that is, the only way that is logically consistent with the most firmly-established principles of fundamental physics) that performance on most applications within realistic power constraints might still continue increasing indefinitely. Reversible logic is also a core part of the quantum circuit model.

An arithmetic logic unit is a multi-functional circuit that conditionally performs one of several possible functions on two operands A and B depending on control inputs. It is nevertheless the main performer of any computing device. The ALU needs to continually perform during the life-time of any computational device such as a computer or a hand held device like PDA (Personal Digital Assistant) etc., Thus heat dissipation becomes a major issue in designing the ALU. Thus reversible logic can be aptly employed in designing the arithmetic logic unit. Also the ALU has to be resistant to the faults that may creep during the operation. Therefore it becomes more suitable that parity preserving reversible logic gates is used to design the ALU.

This paper presents a fault tolerant reversible ALU constructed using parity preserving (also called conservative) logic gates. The rest of the paper is organized as follows: Section 2 gives an overview of reversible logic gates, basic definitions pertaining to them. Section 3 elaborates on the design of ALU using separate arithmetic circuit and logic circuit. Section 4 explains another design of ALU which is implemented using Boolean expressions. Section 5 gives the simulation results and conclusions. Acknowledgements and references follow.

## 2. REVERSIBLE LOGIC

### 2.1. Definitions

Some of the basic definitions [16] pertaining to reversible logic are:

A. Reversible Logic Function: A Boolean Function $f(x_1, x_2, x_3, \ldots x_N)$ is said to be reversible if it satisfies the following criteria : (i)The number of inputs is equal to the number of the number of outputs.(ii)Every output vector has an unique pre-image.
B. Reversible Logic Gate: A reversible logic gate is an N-input N-output logic device that provides one to one mapping between the input and the output. It not only helps us to determine the outputs from the inputs but also helps us to uniquely recover the inputs from the outputs.
C. Garbage: Additional inputs or outputs can be added so as to make the number of inputs and outputs equal whenever necessary. This also refers to the number of outputs which are not used in the synthesis of a given function. In certain cases these become mandatory to achieve reversibility.
D. Quantum Cost: This refers to the cost of the circuit in terms of the cost of a primitive gate. It is computed knowing the number of primitive reversible logic gates (1*1 or 2*2) required to realize the circuit.
E. Gate levels or Logic Depth: This refers to the number of levels in the circuit which are required to realize the given logic functions.
F. Flexibility: This refers to the universality of a reversible logic gate in realizing more functions.
G. Gate count: The number of reversible gates used to realize the function.

A few other cost metrics that have been used for evaluating the performance of reversible logic circuit [11] are the:

   A. Line count-LC- Number of circuit lines in the reversible logic circuit line count directly indicates the number of Qubit (quantum bits) and the circuit cost, a useful parameter for building the quantum circuit.
   B. Transistor Cost: It refers to the number of transistors required if the CMOS technology is adopted for the design.

## 2.2. Basic Reversible Logic Gates

A number of reversible logic gates have been proposed in the literature. The most important among them that are mention worthy are the

**2.1.1. Feynman Gate [5]:** It is a 2x2 gate and its logic circuit is as shown in the figure. It is also known as Controlled Not (CNOT) Gate. It has quantum cost one and is generally used for Fan Out purposes.

**2.2.2. Peres Gate [3]:** It is a 3x3 gate and its logic circuit is as shown in the figure. It has quantum cost four. It is used to realize various Boolean functions such as AND, XOR.

**2.2.3. Fredkin Gate [4]:** It is a 3x3 gate and its logic circuit is as shown in the figure. It has quantum cost five. It can be used to implement a Multiplexer.

**2.2.4. Toffoli Gate [4]:** It is also a 3x3 gate and its logic circuit is as shown in figure. It is a Universal gate. It has a quantum cost of five.

The basic reversible logic gates are shown in the figure 1.

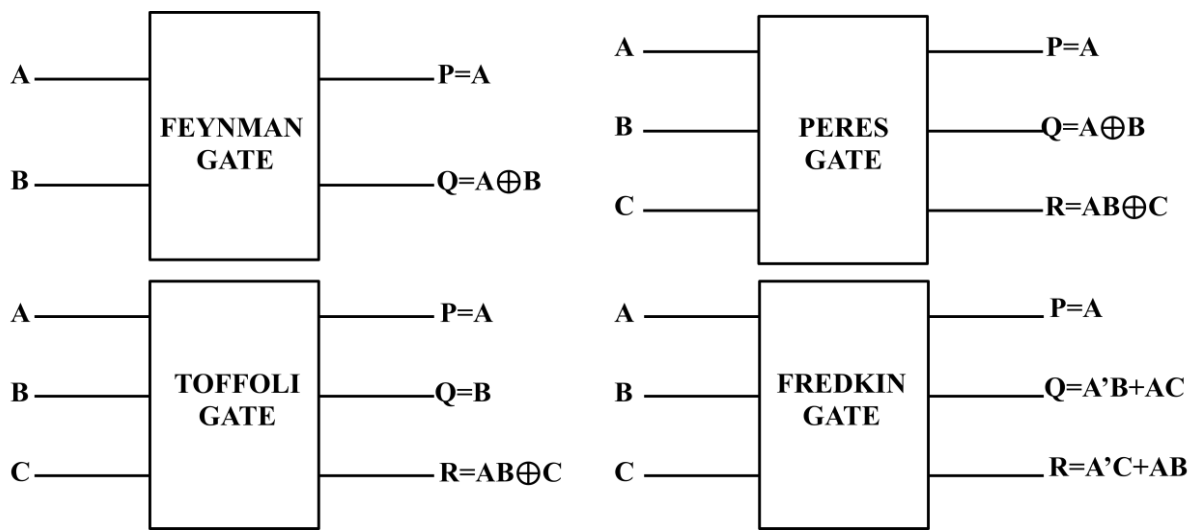

Figure 1. Basic Reversible Logic Gates

## 2.3. Parity Preserving Reversible Logic Gates

The parity preserving reversible logic gates are a class of reversible logic gates with the additional property that the parity of the input is same as the parity of the output. A reversible logic gate will be parity preserving if the EXOR of the inputs matches the EX-OR of the outputs i.e., the parity of the input and the output remains the same. If $I_1, I_2, \ldots, I_N$ and $O_1, O_2, \ldots, O_N$ are the inputs and outputs of an NxN reversible logic gate, it will be parity preserving iff they satisfy
$I_1 \oplus I_2 \oplus \ldots \oplus I_N = O_1 \oplus O_2 \oplus \ldots \oplus O_N$. Some of the parity preserving reversible logic gates that are useful in the discussion are the NFT gate proposed in [7], Islam Gate (IG) [8], Double Feynman Gate

[10], F2PG [13], and PPPG [6]. Also the Fredkin gate is a parity preserving reversible logic gate. All the mentioned gates have been shown in Figure 2.

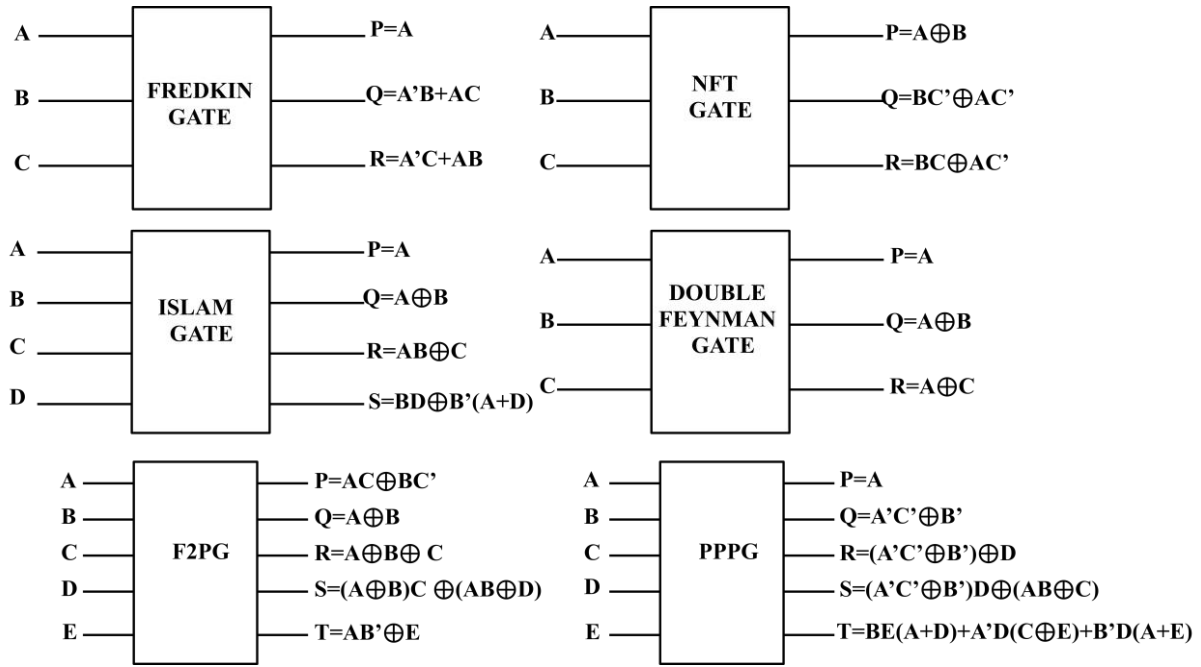

Figure 2. Parity Preserving Reversible Logic Gates

## 3. ARITHMETIC LOGIC UNIT DESIGN 1

### 3.1. Design of Arithmetic Circuit

The basic component of the arithmetic circuit of the ALU is the parallel adder. This is the basic structure, though high speed adders such as Carry Skip Adder, Look Ahead Carry Adder, Carry Save Adder etc., can be used instead of the parallel adder. The parallel adder (also called Ripple Carry Adder) is constructed with a number of full adders connected in cascade. By controlling the data inputs to the parallel adder, it is possible to obtain different types of arithmetic operations. Figure 3 shows one such situation where in input $C_{in}$ is controlled to obtain different operations. The fault tolerant full adders that have been proposed in the literature are in [12], [6], [8] and [13]. These can be used to construct the ripple carry adder. In [12] a generalized structure that can be used to design a fault tolerant full adder is proposed which in turn uses a parity preserving Toffoli gate in conjunction with F2G as shown in figure 4. Then a parity preserving Toffoli structure in [14] is used along with one more proposed structure. In [8] the IG is proposed which is used to build the fault tolerant full adder. In [6] and [13] a 5x5 gate has been proposed namely the PPPG and F2PG which act as a Fault tolerant full adder by applying zeros to two of its inputs. In this case four full adders constitute the parallel adder. The carry into the first stage is the input carry and the carry from the fourth stage is the output carry. All other carries are connected internally from one stage to the next. The selection variables are S0, S1, and $C_{in}$. $S_0$ and $S_1$ control all of the B inputs to the full adder circuits while the A inputs go directly to the other inputs of the full adders.

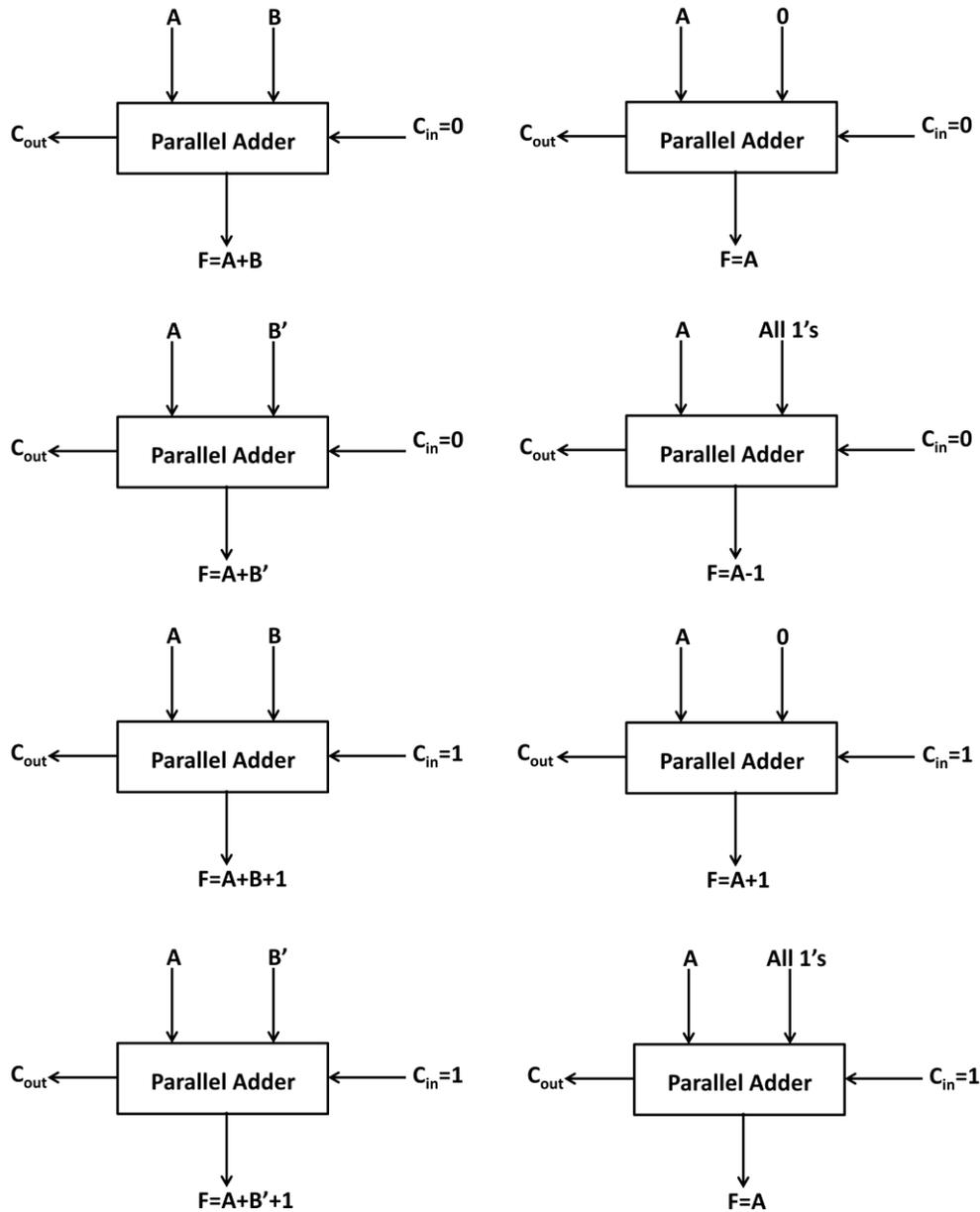

Figure 3. Operations obtained by controlling inputs to parallel adder

The arithmetic operations implemented in the arithmetic circuit are listed in Table 2. The values of Y inputs to the full adder circuits are a function of selection variables S0 and S1. Adding the value of Y in each case to the value of A plus the Cin value gives the arithmetic operation in each entry. The eight operations listed in the table follow from the functional diagrams of Fig. 3. The combinational logic circuit that needs to be inserted before the full adder stage is characterized by the equation:

$$X_i = A_i \qquad (1)$$

$$Y_i = B_i S_0 + B_i' S_1 \qquad i = 1,2,\dots n \qquad (2)$$

Where n is the number of bits in the arithmetic circuit. The reversible implementation of the combinational logic circuit uses a Fredkin Gate to obtain the variable Yi and is as shown in figure 9.

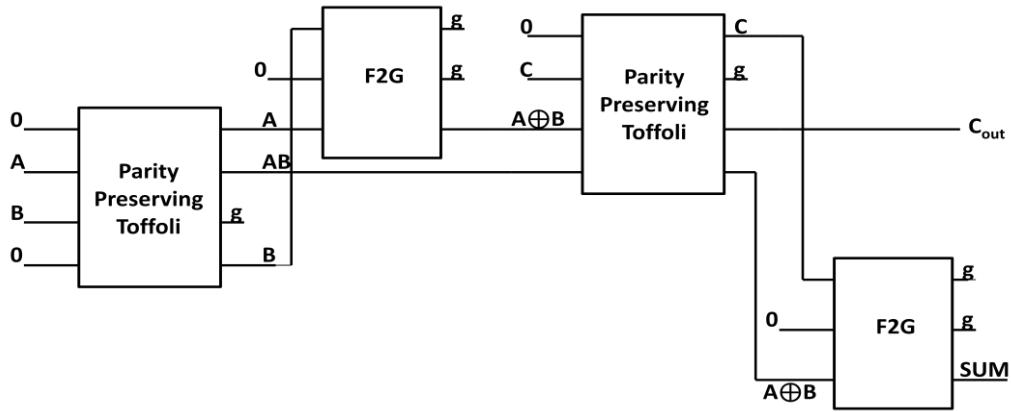

Figure 4. Generalized Structure for Fault Tolerant Full Adder [12]

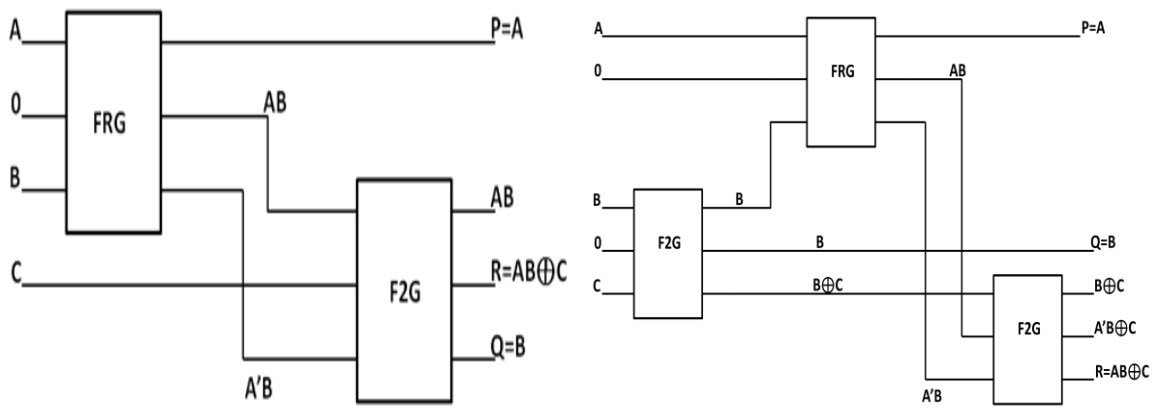

Figure 5. Parity preserving Toffoli Structure in [12] (L) and in [14] (R)

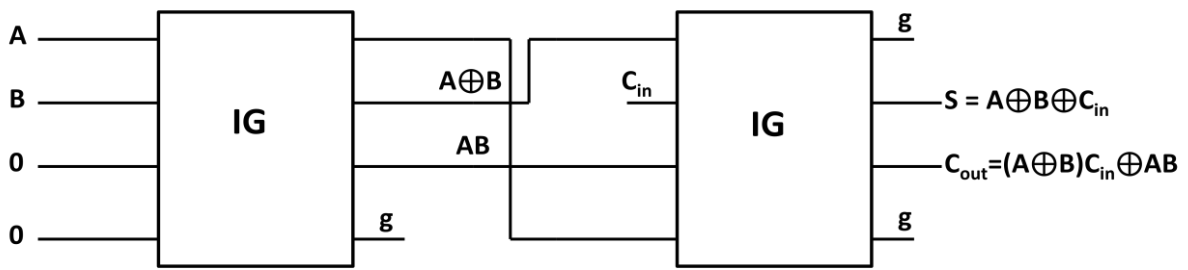

Figure 6. Parity preserving Full adder [8]

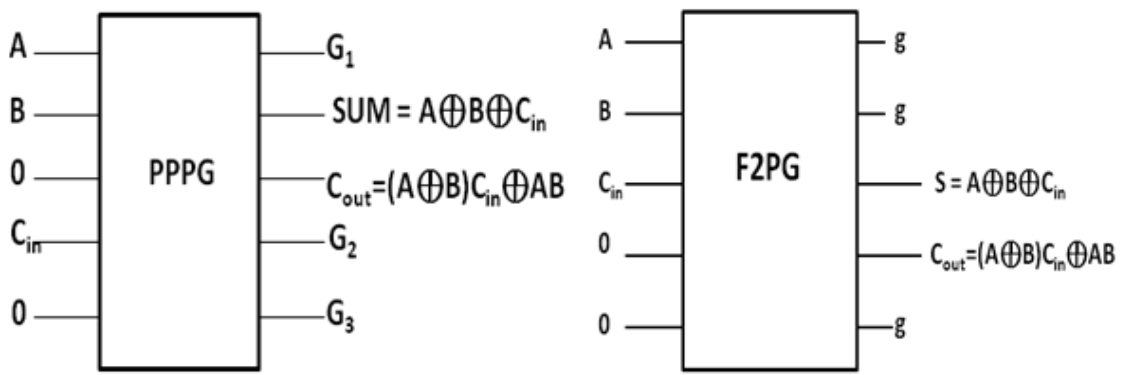

Figure 7. Parity preserving Full adder in [6] and [13]

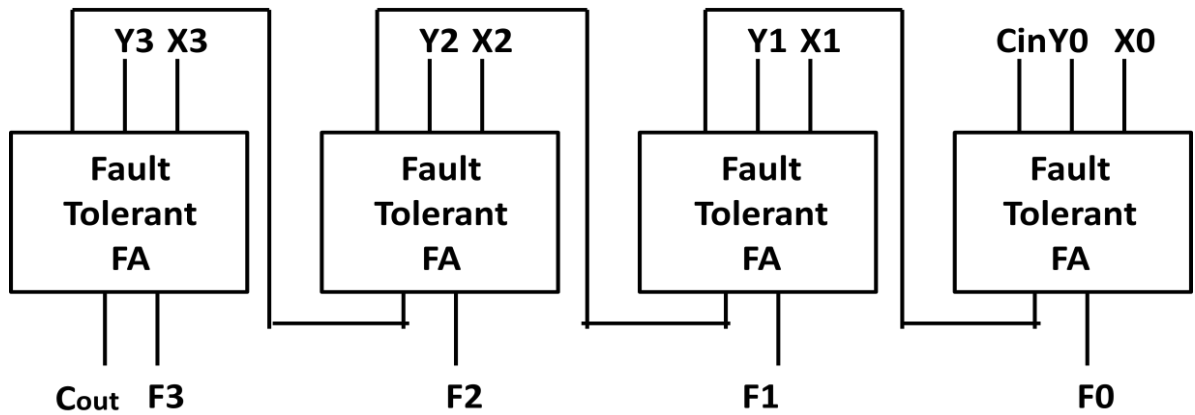

Figure 8. Ripple Carry Adder structure

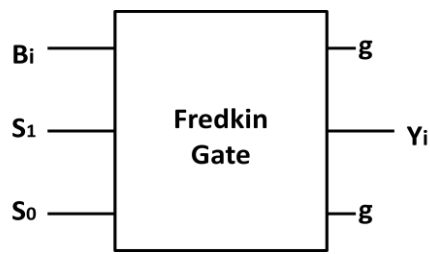

Figure 9. Reversible Implementation of expression (2)

Table 1. Y as a function of $S_1$ and $S_0$

| $S_1$ | $S_0$ | $Y_i$ |
|---|---|---|
| 0 | 0 | 0 |
| 0 | 1 | $B_i$ |
| 1 | 0 | $B_i'$ |
| 1 | 1 | 1 |

The Fault Tolerant Full Adder can be any one of the structures shown in figures 4, 6 or 7. $X_i$ and $Y_i$ given by equations (1) and (2) are implemented using the gate shown in figure 9 and are applied to the ripple carry adder structure in order to generate the output arithmetic functions tabulated in Table 2.

Table 2. Function Table for Arithmetic Circuit

| Function Select | | | Y | Output | Function |
|---|---|---|---|---|---|
| $S_1$ | $S_0$ | $C_{in}$ | | | |
| 0 | 0 | 0 | 0 | $F = A$ | Transfer $A$ |
| 0 | 0 | 1 | 0 | $F = A + 1$ | Increment $A$ |
| 0 | 1 | 0 | B | $F = A + B$ | Add $B$ to $A$ |
| 0 | 1 | 1 | B | $F = A + B + 1$ | Add $B$ to $A$ plus 1 |
| 1 | 0 | 0 | $\overline{B}$ | $F = A + \overline{B}$ | Add 1's Complement of $B$ to $A$ |
| 1 | 0 | 1 | $\overline{B}$ | $F = A + \overline{B} + 1$ | Add 2's Complement of $B$ to $A$ |
| 1 | 1 | 0 | All 1's | $F = A - 1$ | Decrement $A$ |
| 1 | 1 | 1 | All 1's | $F = A$ | Transfer $A$ |

## 3.2. Design of Logic Circuit

The logical unit manipulates the bits of the operands separately and treats each bit as a binary variable i.e., performs bitwise operations. In [15] 16 different logical operations that can be performed on two binary variables have been showcased. These 16 different operations can be generated in one circuit and selected by means of four select lines. Since all logical operations can be obtained by means of AND, OR and NOT operations, it is convenient to employ a logic circuit with just these operations. An XOR function can be chosen to be the fourth operation. One possible way is to implement four of these Boolean functions and select one of them using a 4:1 MUX. The logic diagram and the reversible implementation of the same are as shown in the figure 10 and 11 respectively.

Table 3. Functional Table of Logic Circuit

| $S_1$ | $S_0$ | Output | Operation |
|---|---|---|---|
| 0 | 0 | $F_i = A_i + B_i$ | OR |
| 0 | 1 | $F_i = A_i \oplus B_i$ | XOR |
| 1 | 0 | $F_i = A_i B_i$ | AND |
| 1 | 1 | $F_i = A_i'$ | NOT |

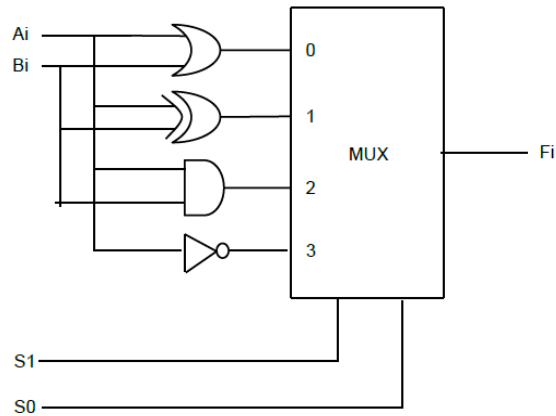

Figure 10. Implementation of Logic Circuit

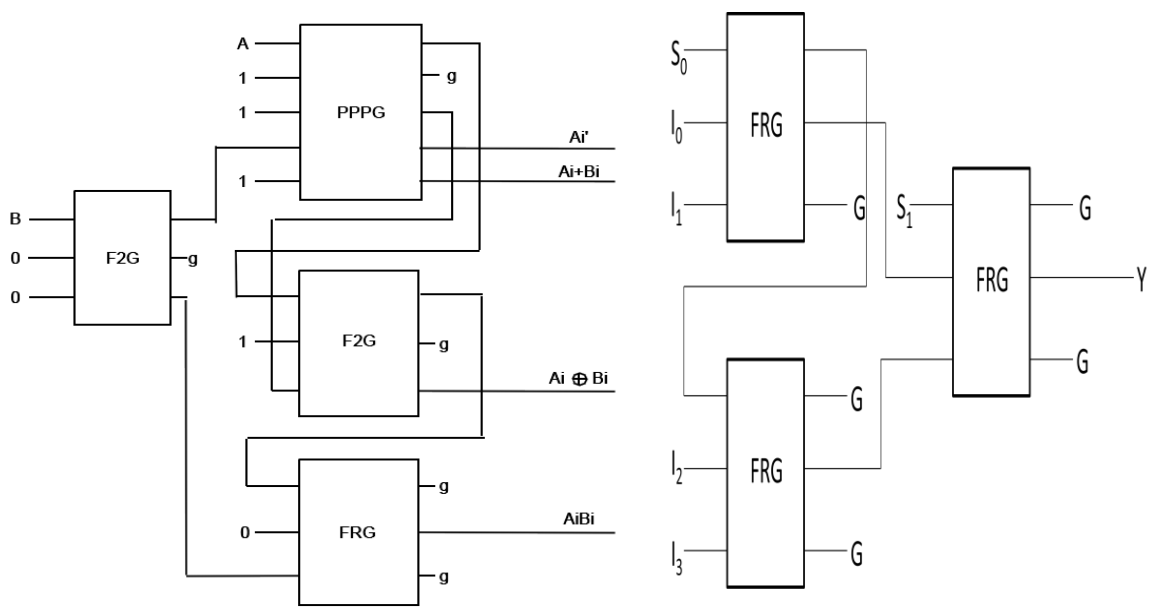

Figure 11. Fault Tolerant Reversible Implementation of Logic Circuit and 4:1 MUX [15]

### 3.3. Design of ALU

The ALU can be constructed by linking the Arithmetic Circuit and the Logic Circuit discussed in the previous two sections. A third select line S3 is used to select either arithmetic or a logic operation at the output of the ALU. Although the two circuits can be combined in this manner, this may not be the best way to design an ALU. The complete block diagram of the one bit ALU is as shown in figure 12. The MUX is implemented using a Fredkin Gate.

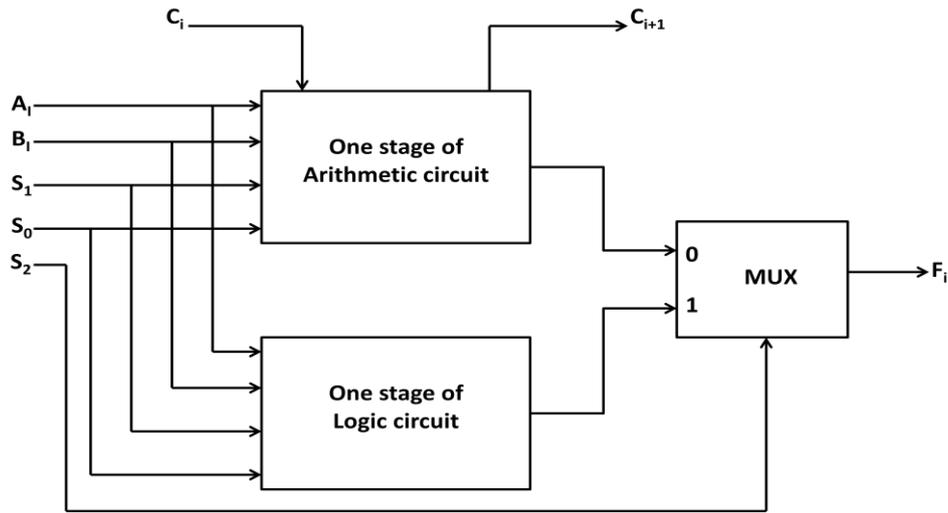

Figure 12. Complete Block Diagram of single Stage ALU.

## 4. ARITHMETIC LOGIC UNIT DESIGN 2

The design of ALU can be done by considering the Boolean Equations and directly implementing them using the parity preserving reversible logic gates. The ALU here also invariably consists of a full adder circuit that is used to generate the final expression. The functional expressions are given by

$$F_i = X_i \oplus Y_i \oplus Z_i \quad (3)$$

$$C_{i+1} = X_i Y_i + Y_i Z_i + X_i Z_i \quad (4)$$

Where $X_i$, $Y_i$, and $Z_i$ are given by expressions

$$X_i = A_i + S_2 \overline{S_0}(S_1 \oplus B_i) \quad (5)$$

$$Y_i = S_0 B_i + S_1 \overline{B_i} \quad (6)$$

$$Z_i = \overline{S_2} C_i \quad (7)$$

When $S_2=0$, the three functions reduce to

$$X_i = A_i$$

$$Y_i = S_0 B_i + S_1 \overline{B_i}$$

$$Z_i = C_i$$

Which are nothing but the functions for the arithmetic circuit. When $S_2=1$, the logical operations are generated as the expressions reduce to

$$X_i = A_i$$

$$Y_i = S_0 B_i + S_1 \overline{B_i}$$

$$Z_i = 0$$

The function generator block takes $A_i$, $B_i$ and $C_i$ as inputs and yields $X_i$, $Y_i$ and $Z_i$ given by equations (5), (6), (7) at the output which are in turn applied to the Full adder in order to generate $F_i$ and $C_{i+1}$ given by equations (3) and (4). The full adder can again be any of the structures shown in figures 4 to 7. The implementation of the function generator done using the parity preserving reversible logic gates is as shown in fig 13.

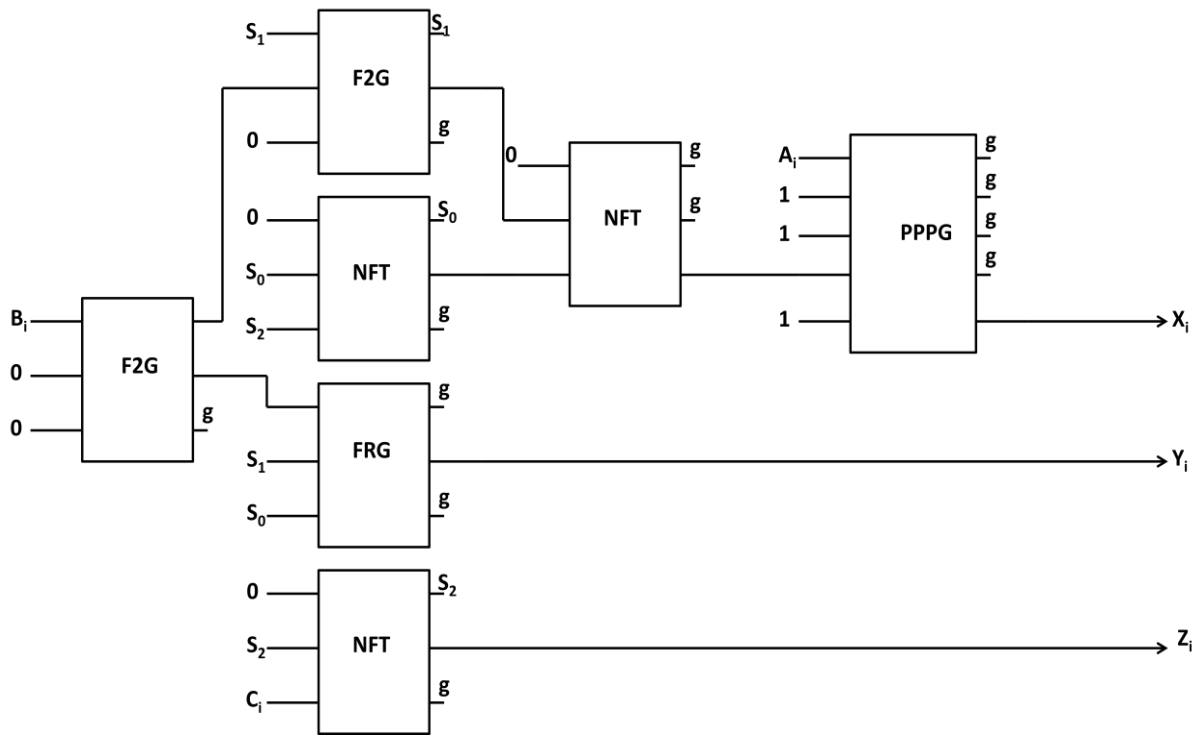

Figure 13. Parity Preserving Function Selector

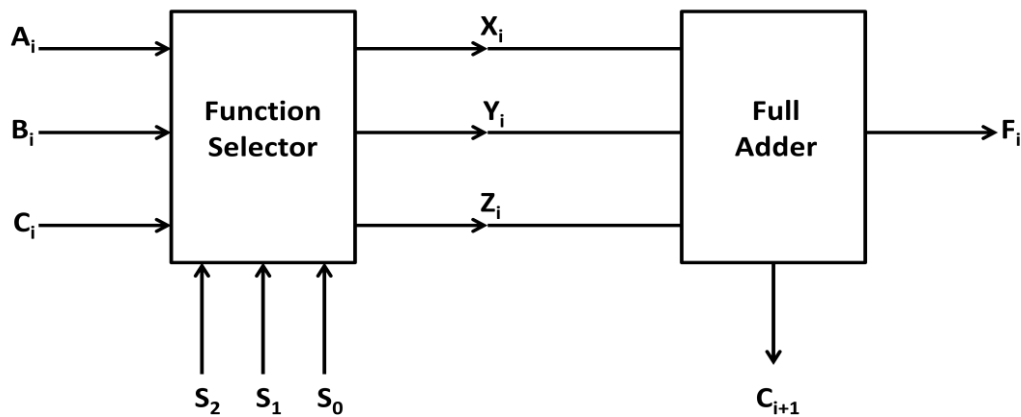

Figure 14. Complete Block Diagram of Optimized Fault Tolerant ALU

The different operations that can be performed by the designed ALU are listed in the form of table 4.

Table 4. Complete Functional Table of the Designed ALU

| Selection | | | | Output | Function |
|---|---|---|---|---|---|
| $S_2$ | $S_1$ | $S_0$ | $C_{in}$ | | |
| 0 | 0 | 0 | 0 | $F = A$ | Transfer $A$ |
| 0 | 0 | 0 | 1 | $F = A + 1$ | Increment $A$ |
| 0 | 0 | 1 | 0 | $F = A + B$ | Addition |
| 0 | 0 | 1 | 1 | $F = A + B + 1$ | Addition with Carry |
| 0 | 1 | 0 | 0 | $F = A - B - 1$ | Subtraction with Borrow |
| 0 | 1 | 0 | 1 | $F = A - B$ | Subtraction |
| 0 | 1 | 1 | 0 | $F = A - 1$ | Decrement $A$ |
| 0 | 1 | 1 | 1 | $F = A$ | Transfer $A$ |
| 1 | 0 | 0 | X | $F = A + B$ | OR |
| 1 | 0 | 1 | X | $F = A \oplus B$ | XOR |
| 1 | 1 | 0 | X | $F = AB$ | AND |
| 1 | 1 | 1 | X | $F = A'$ | NOT |

## RESULTS AND CONCLUSIONS

The arithmetic circuit components viz., the full adder, the ripple carry adder, the logic circuit, the function selector and the complete ALU are all tested for their correctness of logical functionality by simulating them on Xilinx v 9.2i. The simulation results are as shown in figures 15 to 20.

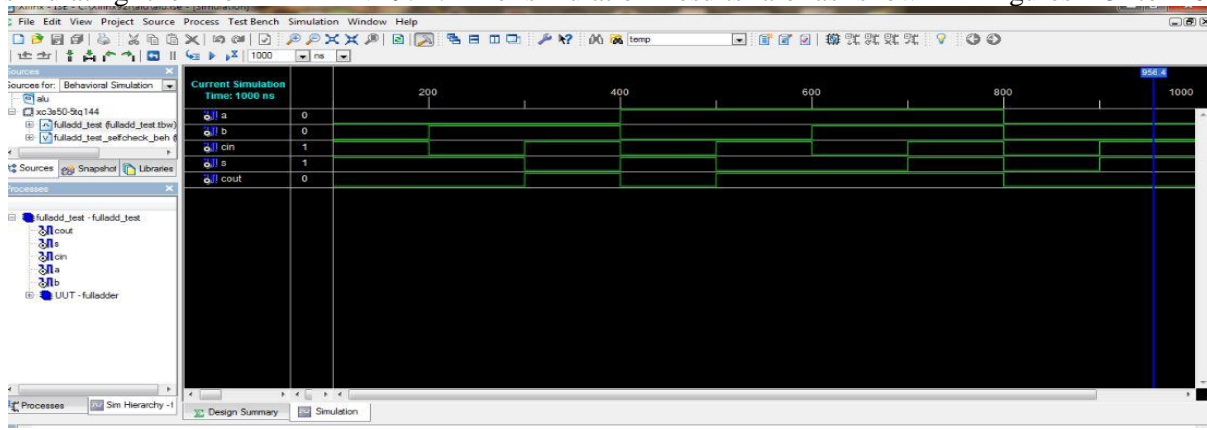

Figure 15. Simulation results of Full Adder Circuit

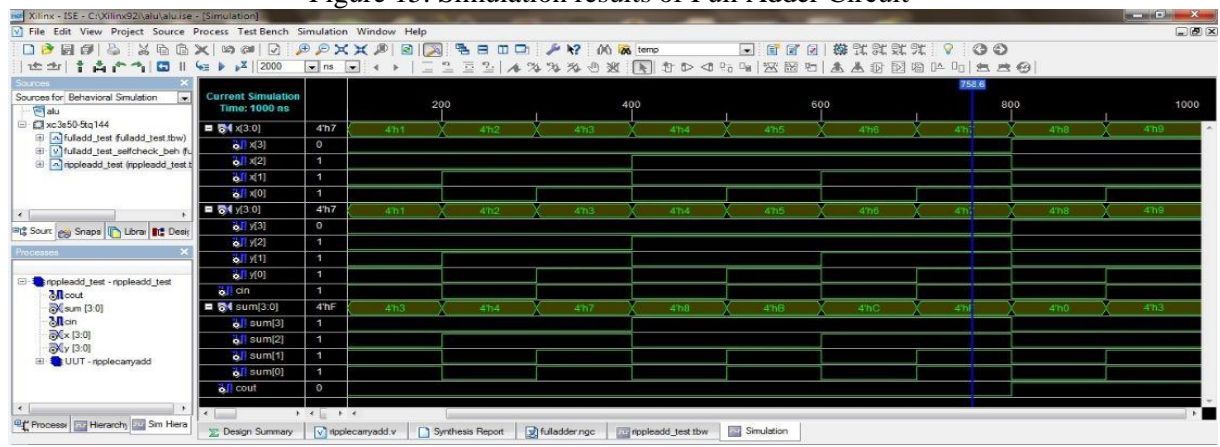

Figure 16. Simulation results of Ripple Carry Adder Circuit along with Arithmetic Unit

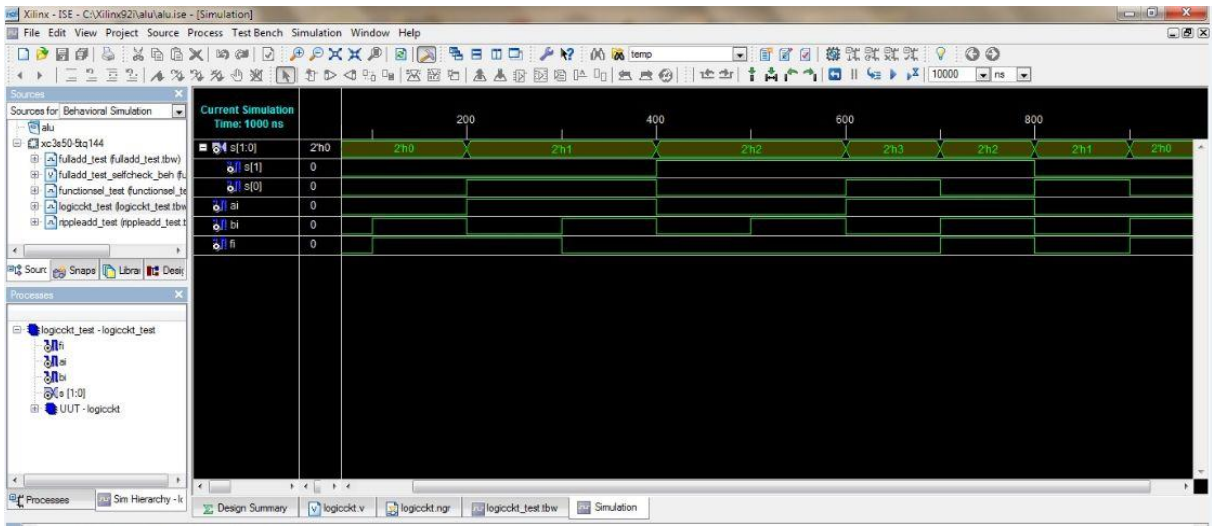

Figure 17. Simulation results of Logic Unit

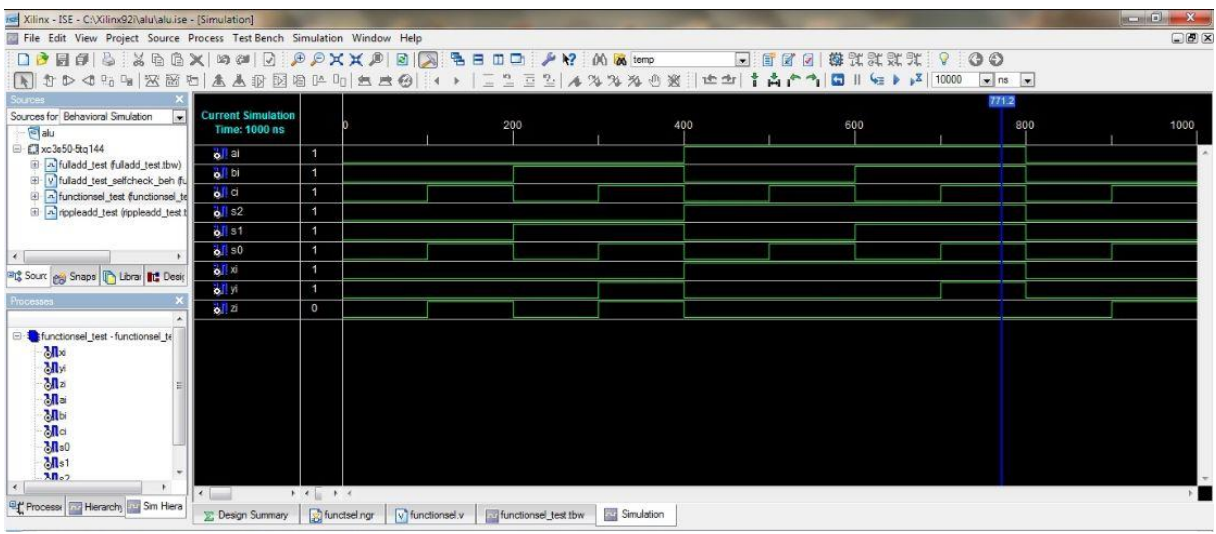

Figure 18. Simulation results of Function Selector

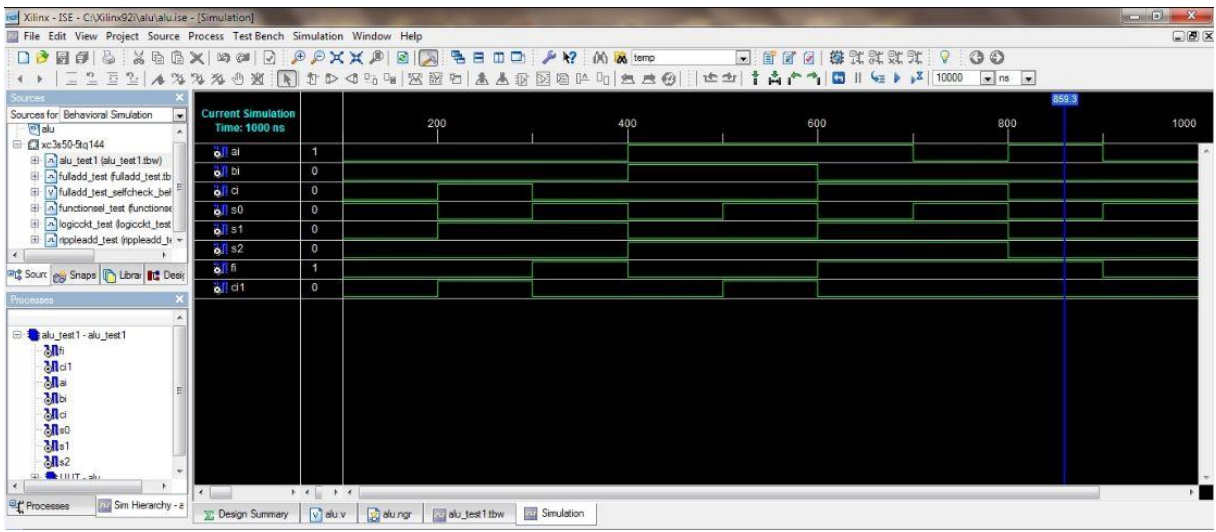

Figure 19. Simulation results of ALU 1

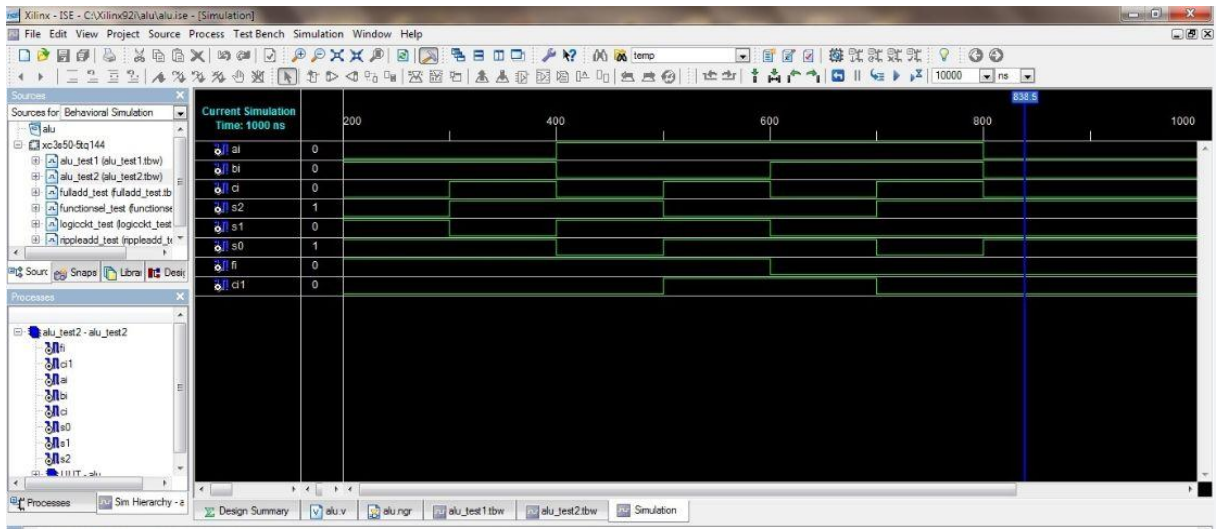

Figure 20. Simulation results of ALU 2

The parameters of the different full adder structure that have been explained in Section 3.1 are tabulated in table 5. The parameters for the arithmetic circuit and the logic circuit are tabulated for a single slice ALU in table 6. Table 7 and Table 8 give the comparison of cost metrics of complete ALU design 1 and design 2. The different cost metrics of the designed ALU can be listed in the form of table as shown below.

Table 5. Comparison of different Full Adder structures

| Full Adder Structure | Gate Count (GC) | Garbage Outputs (GO) | Constant Inputs (CI) |
|---|---|---|---|
| Structure [12] with Toffoli Gate in [14] | 8 | 10 | 9 |
| Structure [12] with Toffoli Gate in [12] | 6 | 8 | 8 |
| [8] | 2 | 3 | 2 |
| [6] | 1 | 3 | 2 |
| [13] | 1 | 3 | 2 |

Table 6. Comparison of Arithmetic and Logic Circuits stemming from Different Full Adder Structures

| Structure | Gate Count (GC) | Garbage Outputs (GO) | Constant Inputs (CI) |
|---|---|---|---|
| Arithmetic Circuit with FA [12]+[14] | 9 | 12 | 9 |
| Arithmetic Circuit with FA [12]+[12] | 7 | 10 | 8 |
| Arithmetic Circuit with FA [8] | 3 | 5 | 2 |
| Arithmetic Circuit with FA [6] | 2 | 5 | 2 |
| Arithmetic Circuit with FA [13] | 2 | 5 | 2 |
| Logic Circuit | 7 | 10 | 7 |

Table 7. Comparison of one slice ALUs in Design 1

| Structure | Gate Count (GC) | Garbage Outputs (GO) | Constant Inputs (CI) |
|---|---|---|---|
| ALU with FA [12]+[14] | 17 | 24 | 16 |
| ALU with FA [12]+[12] | 15 | 22 | 15 |
| ALU with FA [8] | 11 | 17 | 9 |
| ALU with FA [6] | 10 | 17 | 9 |
| ALU with FA [13] | 10 | 17 | 9 |

Table 8. Comparison of one slice ALUs in Design 2

| Structure | Gate Count (GC) | Garbage Outputs (GO) | Constant Inputs (CI) |
|---|---|---|---|
| Function Generator | 7 | 12 | 9 |
| ALU with FA [12]+[14] | 16 | 22 | 18 |
| ALU with FA [12]+[12] | 14 | 22 | 17 |
| ALU with FA [8] | 10 | 17 | 11 |
| ALU with FA [6] | 9 | 17 | 11 |
| ALU with FA [13] | 9 | 17 | 11 |

In this paper, a fault tolerant reversible one slice ALU is constructed and the design can be extended to 4 slice and 8 slice. In design 1, the arithmetic circuit is constructed using full adders and a small combinational logic. The prominent fault tolerant full adder structures in literature have been considered and applied to the design. The logic circuit fundamentally uses a MUX approach as in [15] in order to select one of the 4 logical functions. The arithmetic circuit and logic circuit are then connected to a 2:1 MUX input so as to select an arithmetic function or a logic function at the output of ALU. In design 2, the design equations have been directly implemented using the parity preserving reversible logic gates in order to synthesize a block called function selector that will produce three outputs that are applied to the fault tolerant full adder to obtain the functional ALU. On the concluding lines, reversible logic is one of the emerging computing paradigms that have potential for generating zero power dissipation and an ALU being the heart of any processor, the reversible implementation of the same using reversible logic is bound to have a major impact on nanotechnology based systems.

## ACKNOWLEDGEMENTS

The authors would like to thank Department of Electronics and Communication, B.M.S. College of Engineering, Bangalore India for supporting this work.

## REFERENCES

[1] R. Landauer,"Irreversibility and Heat Generation in the Computational Process", IBM Journal of R&D,1961
[2] C.H. Bennett, "Logical reversibility of Computation", IBM J. Research and Development, pp.525-532, November 1973.
[3] A. Peres, Reversible logic and quantum computers, Phys. Rev. A 32 (1985) 3266–3276.
[4] E. Fredkin and T. Toffoli,"Conservative Logic", Int'l J. Theoretical Physics Vol 21, pp.219-253, 1982.
[5] R Feynman " Quantum Mechanical Computers", Optical News, Vol.11, pp 11-20, 1985
[6] Krishna Murthy, Gayatri G, Manoj Kumar "Design of Efficient Adder Circuits Using Proposed Parity Preserving Gate" VLSICS Vol.3, No.3, June 2012.
[7] Haghparast, M. and K. Navi, " A novel fault tolerant reversible gate for nanotechnology based systems". Am. J. Appl. Sci., 5(5).2008

- [8] Md. Saiful Islam et.al" Synthesis of fault tolerant Reversible logic"IEEE 2009
- [9] Rakshith Saligram and Rakshith T.R. "Design of Reversible Multipliers for linear filtering Applications in DSP" International Journal of VLSI Design and Communication systems, Dec-12
- [10] B. Parhami, "Fault tolerant reversible circuits", Asimolar Conf. Signal systems and computers", October 2006
- [11] H R Bhagyalakshmi and M K Venkatesha," Optimized multiplier using Reversible Multi-Control Input Toffoli Gates", VLSICS, Vol 3. No (6), Dec.- 12
- [12] Majid Haghparast and Keivan Navi, "Design of a Novel Fault Tolerant Reversible Full Adder for Nanotechnology Based Systems", World Applied Sciences Journal 3 (1): 114-118, 2008.
- [13] Xuemei Qi and Et. Al, Design of fast fault tolerant reversible signed multiplier, International Journal of the Physical Sciences Vol. 7(17), pp. 2506 - 2514, 23 April, 2012.
- [14] Parhami, B., 2006. Fault tolerant reversible circuits, Proc. 40th Asilomar Conf. Signals, Systems and Computers, October 2006, Pacific Grove, CA, pp: 1726-1729.
- [15] Rakshith Saligram, Shrihari S.H., Shashidhar A.K. H.R. Bhagyalakshmi, M.K. Venkatesha, "Design of Fault Tolerant Reversible Multiplexer Based Multi-Boolean Function Generator using Parity Preserving Gates", International Journal of Computer Applications, March 2013.
- [16] Rakshith Saligram, Rakshith T R,"Novel code converters employing reversible logic", International Journal of Computer Applications, Vol. 52 Aug 2012.
- [17] Michael P. Frank, Introduction to Reversible Computing: Motivation, Progress, and Challenges.